# Contactless determination and parametrization of charge carrier mobility in silicon as a function of injection level and temperature using time resolved THz spectroscopy

*Sergio Revuelta and Enrique Cánovas**

IMDEA Nanociencia, Campus Universitario de Cantoblanco, Faraday 9, 28049 Madrid, Spain

* enrique.canovas@imdea.org

**Abstract**

Here, we analyze in a non-contact fashion charge carrier mobility as a function of injection level and temperature in silicon by time resolved THz spectroscopy (TRTS) and parametrize our data by the classical semi-empirical models of Klaassen and Dorkel & Leturcq. Our experimental results are in very good agreement with the pioneering works of Krausse and Dännhauser analyzing this phenomena by employing contact-based methods. This agreement, that validates our methodology, can only be achieved by considering charge carrier diffusion effects following above bandgap near-surface pump photo-excitation of the sample. From our results, obtained over a large range of injection levels, we conclude that the model of Klaassen is the best on describing the collected data at room temperature. Furthermore, we analyze by TRTS the dependence of charge carrier mobility with temperature for a fixed injection level. Once more, the parametrization made by the classical model of Klaassen describe our data appropriately even without the necessity of applying any fitting parameters (just with the charge carrier density as an input). In this respect, our work supports the validity of the model and parametrization proposed by Klaassen, and also illustrate how TRTS can be reliably employed for the quantitative determination of mobility in semiconductors as a function of key parameters as injection level and temperature.

Charge carrier mobility is a fundamental figure of merit determining the performance of opto-electronic devices. This variable depends critically on few parameters as the charge carrier concentration via doping, the temperature and the injection level [1,2]. To our knowledge, pioneering studies of the charge-carrier mobility dependence vs injection level in silicon were made in the 70s from contact-based techniques [3,4] and these results still represents nowadays the basis for semi-empirical models employed in modelling charge carrier mobility as a function of charge carrier density in silicon [5–9]. These initial studies analyzed the interplay of mobility and injection level by contact-based approaches, and hence required the manufacturing of a fully functional working device (e.g., a PIN or PN diode structure).

Over the years all-optical non-contact methods have been employed as alternatives for scrutinizing the charge carrier mobility-charge carrier injection relationship, most notably photoconductance decay (PCD) measurements. This technique is very powerful for estimating the mobility, recombination processes and carrier lifetimes in semiconductors as a function of injection level, however it cannot disentangle (from conductance) the contributions of carrier density and mobility [10–13] unless it is assisted by a second set of measurements, e.g. time-resolved photoluminescence (TRPL) [13]. Furthermore, PCD techniques are often able to monitor a relatively narrow range of injection levels (i.e. charge carrier densities). A powerful non-contact alternative to PCD is Time-Resolved THz spectroscopy (TRTS) [14–16]. Following an optical pump-THz probe scheme it is possible to retrieve the frequency-resolved complex photoconductivity of a given sample at any pump probe delay, from which mobility and carrier density can be independently inferred. A TRTS study as a function of impinging - above band gap - pump photon flux, and hence injection level, offers an avenue for determining the interplay between photo-generated charge carrier density and charge carrier mobility. Although several studies have been already done analyzing by TRTS the mobility dependence on charge carrier density in materials as silicon [17,18], gallium arsenide [19,20], titanium oxide [18] or zinc oxide [21], to our knowledge none of these studies have attempted, nor reached a quantitative agreement with classical estimates made by contact methods; a singular aspect that represents by itself a goal between the THz and metrology research communities [22].

In this work, we have employed TRTS to analyze the dependence of sample's charge-carrier mobility with injection level and temperature as controlled by photo-doping in silicon. From TRTS data at room temperature, we retrieve the expected reduction in the charge carrier mobility as a function of injection level in a trend that closely matches the one obtained in classical works employing contact-based methods. Notably, this agreement can only be reached by considering diffusion effects impacting the TRTS data and analysis, which, we show here can critically play a role on determining unambiguously bulk charge carrier densities under our TRTS conventional experimental conditions (i.e. near surface photoexcitation with UV photons). Furthermore, we have analyzed the dependence of charge carrier mobility with temperature at low injection levels and compared our data to the predictions made by the classical semi-empirical models of Klaassen [6,7] and Dorkel & Leturcq [5]. Our results, retrieved over a large range of injection levels and temperatures reveal that the model of Klaassen is the best on describing the data at room temperature and as a function of temperature. For the latter variable even without the necessity of applying any fitting parameters (just the charge carrier density as an input). In this respect, our work supports the validity of the model and parametrization proposed by Klaassen, and also demonstrates that TRTS can be employed as powerful tool for the quantitative determination of mobility in semiconductors, an aspect validated by the good agreement found between our data and the classical results made by Krausse and Dännhauser by contact methods [3,4].

I. EXPERIMENTAL

The sample analysed in this work consisted of a 0.5mm thick semi-insulating silicon float zone (FZ-Si) wafer with <100> orientation (Sigma-Aldrich ID: 646687, resistivity 100-3000 Ω·cm) with native oxide passivation. A Ti:Sapphire amplified laser system providing 775 nm wavelength output (~150 fs pulse width at 1 KHz repetition rate) was employed to run the optical pump-THz probe experiment [14,18]. For optical pump excitation of the sample, we employed the 387.5 nm output generated by a Beta Barium Borate (BBO) crystal. The employed ~1THz bandwidth probe was generated via optical rectification on a 1 mm thick ZnTe crystal cut along the <110> axis. The detection of the THz beam was performed via electro-optical sampling on a ZnTe crystal of identical characteristics.

In a TRTS measurement we are able to monitor the ~1THz width freely propagating probe in the time domain, as such, changes in amplitude and phase induced in the transmitted THZ probe by the pump excitation can be recorded. From this data, the time dependent sample´s sheet frequency resolved complex photoconductivity, $\Delta\sigma_{sheet}(\omega,t)$ can be retrieved at any pump-probe delay time ($t$). Under the employed photo-excitation conditions (387.5 nm) the optical penetration depth in silicon is estimated to be ~80 nm [23], much smaller than the sample´s thickness. This allows us to employ the Tinkham approximation [14,23] from which we can retrieve the frequency resolved complex photoconductivity as:

$$\Delta\sigma_{sheet}(\omega,t) = \frac{-(n_1+n_2)}{Z_0}\frac{\Delta E(\omega,t)}{E_{ref}(\omega)} \qquad \text{(eq. 1)}$$

Where $Z_0$ is the intrinsic impendance of free space, $n_1$ is the refractive index of the medium before the sample and $n_2$ is the refractive index of the photo-excited material, $E_{ref}$ is the transmitted THz electric field through the unexcited sample and $\Delta E$ is the pump-induced change in the THz waveform.

## II. RESULTS

In figure 1a we show the frequency resolved complex sheet photoconductivity obtained for silicon for three pump-probe delays (from top to bottom: 60, 460 and 1260 ps) at a fluence of 10.58 µJ/cm² at 387.5 nm of pump wavelength. Black solid and blue open circles are the experimental real and imaginary components of the sheet complex conductivity within the probed THz window. Each plot also shows black solid and blue dashed lines representing respectively the best fits to real and imaginary components of the frequency resolved complex conductivity by using the Drude model as:

$$\Delta\hat{\sigma}_{sheet}(\omega,t) = \Delta\hat{\sigma}_{bulk}(\omega,t)\cdot d = \frac{N_{sheet}e^2\tau}{m^*(1-i\omega\tau)} \qquad \text{(eq. 2)}$$

Here, $\Delta\hat{\sigma}_{bulk}(\omega,t)$ represents the complex frequency resolved bulk photoconductivity, $d$ is the penetration depth of the impinging photon flux, $e$ is the electron charge, $\tau$ is the average scattering time, $m^*$ is the effective mass and $N_{sheet}$ is the sheet charge carrier density in units of cm$^{-2}$ (i.e. $N_{sheet} = [1-R]N_{h\nu}$; where $R$ and $N_{h\nu}$ are the reflectivity and pump photon flux respectively). From the best Drude fits to the data we can directly obtain values for the sheet carrier density (by considering an optical effective mass of $0.16 m_e$ [8]), $N_{sheet}$, of 7.97·10$^{16}$, 7.55·10$^{16}$ and 7.08·10$^{16}$ m$^{-2}$ and averaged scattering times, $\tau$, of 95, 123 and 141 fs respectively for the analysed pump-probe delays of 60, 460 and 1260 ps. These figures show an improvement in scattering rate upon pump-probe delay that has been previously interpreted by some authors as a diffusion process of photo-generated charge carriers from near the surface towards the bulk [20]. In this respect, an increase in scattering rate is linked with a reduction in bulk charge carrier density of the electron-hole plasma as modulated by a time dependent penetration depth. Then, $d$ in eq. 2 should be defined as $d_{eff}(t)$ (see Figure 1b where $d_{eff}(t)$ represents the distance from the sample´s surface where the initial carried density amplitude drops a value of $1/e$).

Right after charge carrier photogeneration ($d_{eff}(t\sim0)$), the penetration depth for the photo-excited slab in silicon is estimated to be ~80 nm [24] under the 387.5 nm above bandgap pump beam. As neither hot carrier cooling [25], nor charge carrier surface/bulk recombination substantially affect our dynamics within the analysed ~1ns probed time window (see Fig. S2 in Supplemental Material [26]), we can find a numerical solution for the effective time dependent thickness $d_{eff}(t)$ of the expanding electron-hole plasma following the expression $n(x,t) = \left(N_0/\sqrt{4\pi D_{ab} t}\right) \cdot e^{-x^2/4D_{ab} t}$; where $N_0$ refers to the initial carrier density and $D_{ab}$ is the ambipolar diffusivity in silicon [27,28]. The ambipolar diffusivity is parametrized with the scattering time of the electron-hole plasma retrieved from the frequency-resolved complex sheet photoconductivity (see eq. S2 in Supplemental Material [26]).

Once the effective penetration depth is accurately estimated, we can obtain the sample's bulk photoconductivity as $\Delta\hat{\sigma}(\omega,t) = \Delta\hat{\sigma}_{sheet}(\omega,t)/d_{eff}(t)$. Analogously, we can obtain the bulk carrier density as a function of time as $N_{bulk}(t) = N_{sheet}/d_{eff}(t)$, which corresponds to the photo-injected density of charge-carriers. From the results shown in Figure 1a, we retrieve charge carrier concentrations of *(2.0±0.3)·10$^{17}$, (6.6±1.4)·10$^{16}$ and (3.5+0.5)·10$^{16}$* cm$^{-3}$ and charge carrier mobilities of 640 ± 22, 831 ± 43 and 956 ± 38 cm$^2$V$^{-1}$s$^{-1}$ for the pump-probe delays of 60, 460 and 1260 ps respectively (notice here that mobility is defined as $\mu[N_{bulk}] = e \cdot \tau[N_{bulk}]/m^*$).

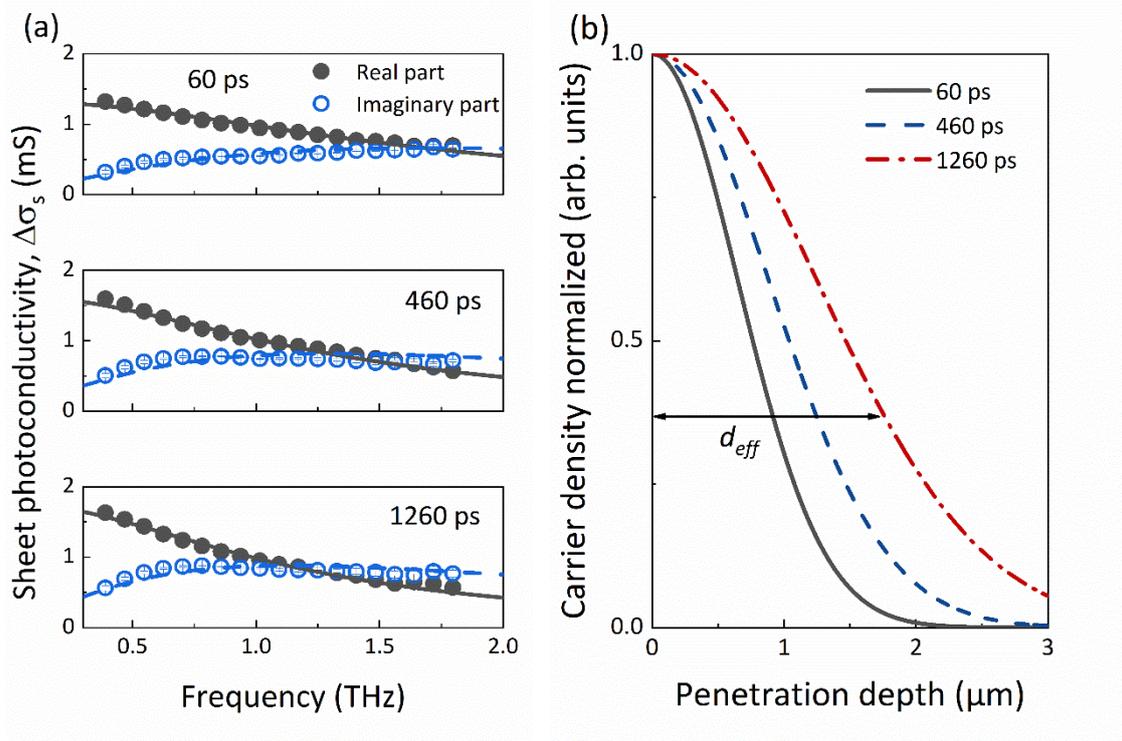

**Figure 1.** (a) Complex sheet photoconductivity at 300 K and under 387.5 nm of pump excitation with a fluence of 10.58 µJ/cm$^2$ at different pump-probe delays (60, 460 and 1260 ps). Filled and open symbols refer to the real and imaginary components of the sheet photoconductivity. Solid and dashed lines correspond to the best fit by the Drude model for the real and imaginary components respectively. (b) Numerical solution for the effective thickness of the photo-generated electron-hole gas for the considered pump-probe delays. The effective thickness for the photo-excited slab is defined as the 1/e drop from the carrier density near the air-solid interface.

In order to infer the variation of charge carrier mobility vs injection level, we performed TRTS experiments for a broad range of 387.5 nm pump excitation conditions ranging between ~0.15 and ~70 µJ/cm$^2$ and at pump-probe delays higher than 10 ps, thereby ensuring the complete cooling of hot carriers [25] (see Fig. S3 in Supplemental Material [26]). The employed upper threshold photon flux was simply limited by our set-up specs. For small photo-excitation conditions (below ~1 µJ/cm$^2$) the recorded frequency resolved complex photoconductivity was found to be invariant vs photon flux and pump-probe delay. Beyond this threshold, an increase in photon flux alone for a given pump-probe delay was sufficient to modulate the charge carrier scattering rate, in good agreement with a previous TRTS study made in silicon [18]. Furthermore, for each given fluence above the mentioned threshold of ~1 µJ/cm$^2$, we found the same temporal trend shown in figure 1, i.e., an improvement in charge carrier scattering rate as a function of pump-probe delay. Taking into account diffusion processes for each analysed fluence (and pump-probe delay), an accurate bulk carrier density can be inferred and linked to the obtained scattering rate from TRTS analysis. In agreement with a previous study conducted in GaAs which considered diffusion effects following photo-excitation [20], we found a substantial overlap between the data obtained from different optical pump-THz probe measurements made at different pump fluences (Figure 2a and Table S3 in Supplemental Material [26]), an aspect that supports the validity for the diffusion model depicted in Figure 1b. Figure 2b summarizes our findings as open diamonds regarding charge carrier mobility vs injection level. From the figure, it is clear that injection level modulates charge carrier mobility in the sample for carrier concentrations above ~10$^{16}$ cm$^{-3}$. In Figure 2b we also include as solid circles the classical data obtained in the 70s by Krausse and Dännhauser [3,4] representing the sum mobility ($\mu_e + \mu_h$) as a function of injection level measured by contact-based methods on working devices. Remarkably, the optically retrieved THz mobility as function of photo-induced carrier density agrees quite well with the electrical carrier mobility obtained via contact methods. To our knowledge this is the first time such correlation is made by TRTS, and over such a large range of injection levels (higher than 1·10$^{16}$ cm$^{-3}$) when compared to e.g. photoconductance based non-contact methods [13,29]. Here, we warn the reader that this agreement between TRTS data and the one obtained by contact methods is only reached if diffusion of photogenerated charge carriers from the surface towards the bulk is considered, otherwise bulk charge carrier densities are strongly overestimated (see

Supplemental Material, Figure S4 [26]).

In order to parametrize the observed experimental results summarized in Figure 2b we fit the data by the classical models developed by Dorkel & Leturcq [5] (dashed line) and Klaassen (solid line) [6,7]. In brief, Dorkel & Leturcq approach considers the mobility contributed by phonon-carrier, impurity-carrier and carrier-carrier interactions. On the other hand, Klaassen model employs a more refined model by considering phonon-carrier, electron-hole, effects of the majority and minority impurity scattering while including a screening effect and an increase of the system temperature with charge carrier density. A summary and of the parameters employed in the models can be found in Table S1 and Table S2 in the Supplemental Material [26]. From our results, we can conclude that the Klaassen model offers a better description of the experimental data retrieved by TRTS. A conclusion that agrees with non-contact PCD data collected over a narrower injection level window [9,13,29,30]. At present we cannot rationalize the slight deviation between TRTS data and those retrieved by contact methods for concentrations above $1\cdot10^{16}$ cm$^{-3}$, in this respect, Klassen suggests that this data might be underestimated by an increase of temperature in the system during the measurements [6].

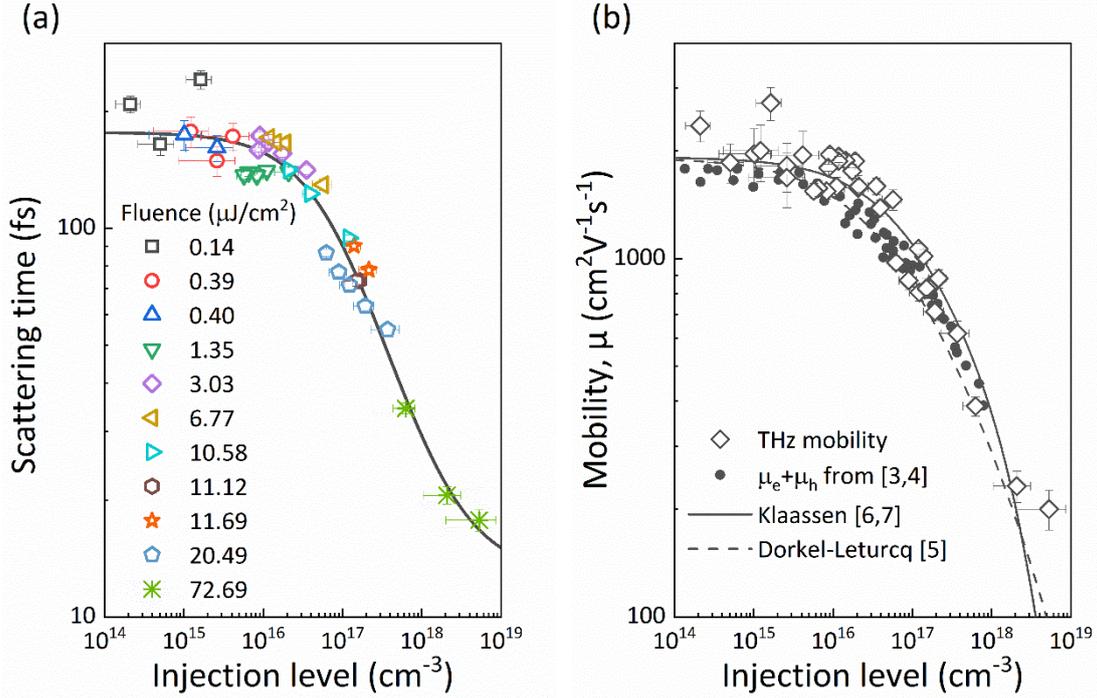

**Figure 2.** (a) Scattering time as function of injection level. Different symbols correspond to different fluences as indicated. Solid line is the best fit to a hyperbolic function as described in the text. (b) Room temperature charge-carrier mobility as function of injection level. Open diamonds refer to experimental points retrieved from TRTS. Black solid dots correspond to the charge-carrier mobility measured by Krausse and Dännhauser [3,4]. Solid dashed line indicates the theoretical value predicted by Dorkel & Leturcq [5] while solid line corresponds to the model described by Klaassen [6,7].

Once we have analysed the dependence of the mobility vs injection level by TRTS at room temperature, we study the dependence of the mobility as a function of temperature for a fixed sheet charge carrier injection. To study this dependence, we selected a photon flux of ~1.55 µJ/cm$^2$; low enough to prevent strong diffusion effects in the samples. Figure 3 shows exemplary measurements of the frequency resolved complex sheet photoconductivity under 387.5 nm excitation at 1 ns of pump-probe delay and different temperatures: 90, 150, 220 and 300 K. All the plots can be properly described by the Drude model. Considering a temperature dependent ambipolar diffusivity (see Figure S1 in Supplemental Material [26]) we estimate charge carrier densities of $(1.98\pm0.13)\cdot10^{16}$, $(1.88\pm0.23)\cdot10^{16}$, $(1.57\pm0.27)\cdot10^{16}$ and $(1.44\pm0.21)\cdot10^{16}$ cm$^-$

[3] which are linked to charge carrier mobilities of 5149 ± 373, 3499 ± 373, 2501 ± 322 and 1735 ± 164 $cm^2V^{-1}s^{-1}$ respectively for the above-mentioned temperatures. As anticipated by the selection of the low photon flux, the carrier concentration values are almost identical as a function of temperature, ranging from ~$2·10^{16}$ to ~$1.4·10^{16}$ $cm^{-3}$, this aspect is critical for parametrizing the collected mobility data for a fixed carrier concentration. In this sense, the obtained variation of mobility vs T, truly represents the impact of temperature on the monitored scattering rates for a given charge carrier injection. Figure 3b summarizes, as open diamonds, the charge carrier mobility inferred from TRTS analysis for a set of temperatures ranging between 77 and 300 K (1.55 µJ/$cm^2$, 387.5 nm and 1 ns of pump-probe delay, see Figure S5 in Supplemental Material for the rest of the frequency-resolved data [26]). Figure 3b also includes a parametrization of the obtained data following the models of Klaassen [6,7] (solid line) and Dorkel & Leturcq [5] (dashed lines); where remarkably the only variable input is the charge carrier concentration figure of $1.73·10^{16}$ $cm^{-3}$ (i.e. the median value of the retrieved carrier densities from the Drude fits). It is evident form the plot that the mobility inferred by TRTS matches nicely the curve predicted by Klaassen [6,7] while the Dorkel & Leturcq [5] model is unable to correctly describe the experimental data.

The validation made here for the Klaassen model as a function of temperature for a given injection level is to our knowledge unique. We are only aware of a couple of PCD works that have previously attempted such study. In these works they found that Klaassen model failed to describe the experimentally resolved trend for temperatures below ~150 K [9,29]. The reason for this disagreement is unclear to us, however, we note that for low temperatures, one should consider an extra effect decreasing charge carrier density in silicon, an effect linked to the condensation of photogenerated e-h pairs into excitons. This effect indeed has been previously reported and modelled to occur at temperatures around the ~100 K onset [31–33].

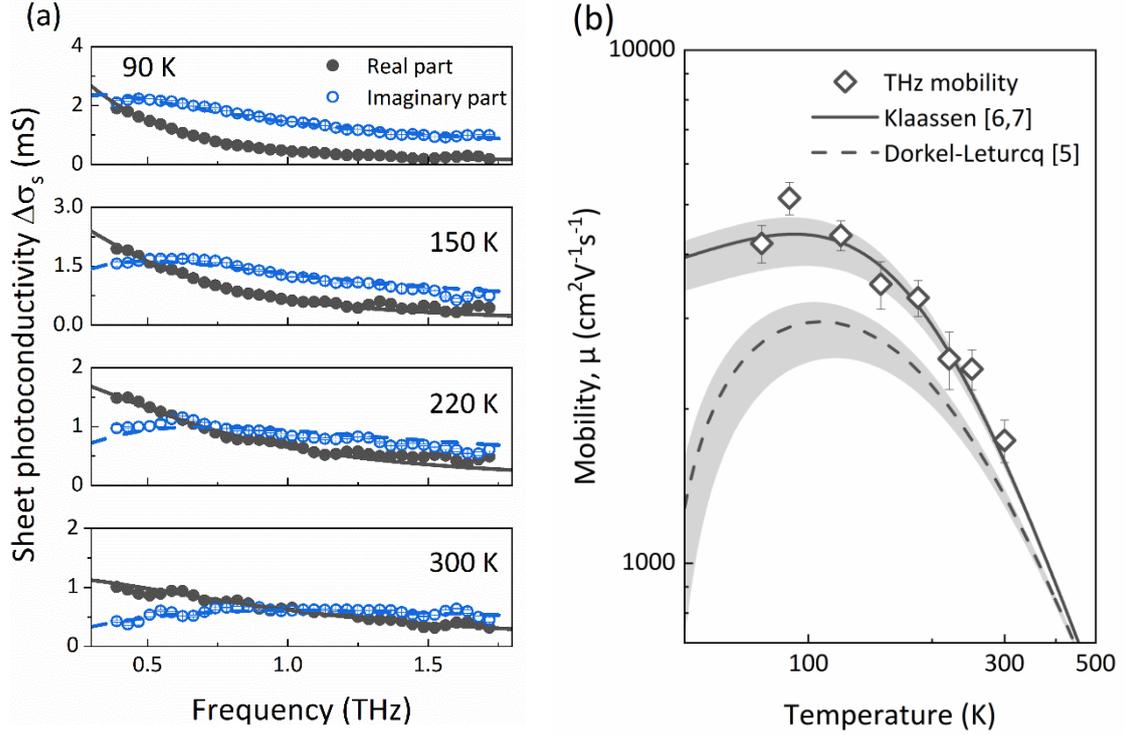

**Figure 3.** (a) Complex sheet photoconductivity for silicon measured with different temperatures at a fluence of 1.55 µJ/cm$^2$ and fixed pump-probe delay of 1 ns. Filled and open symbols refer to the real and imaginary components of the sheet photoconductivity. Solid and dashed lines represent the best fit of the Drude model for the real and imaginary parts respectively. (b) Temperature dependence on charge-carrier mobility retrieved from the signatures of the frequency-resolved photoconductivity. Open black symbols refer to experimental points while the solid dashed line indicates the carrier mobility predicted by Dorkel & Leturcq [5] and the solid line corresponds to the model described by Klaassen [6,7]. The grey areas represent the lower and maximum error deviations on the avg inferred charge-carrier density within the analyzed range of temperatures.

## III. CONCLUSSIONS

In this work we have investigated the dependence of the charge carrier mobility vs injection density and temperature in silicon using time resolved THz spectroscopy. Our results agree well with previously reported studies in silicon measured by contact methods [3,4] and validate the semi-empirical model developed by Klaassen [6,7].

Notably, the agreement with previous data and modelling can only be achieved when charge carrier diffusion effects following near surface photo-excitation are considered. Otherwise, the charge carrier density may be over-estimated for a given charge carrier mobility. The retrieved dependency of the carrier mobility for a fixed injection level with temperature further supports the validity of the Klaassen model against the one proposed by Dorkel & Leturcq.

While in principle our approach can be generalized to any semiconductor, surface recombination effects might complicate or make not possible determining unambiguously charge carrier density as a function of time after excitation. Furthermore, a pre-knowledge of the diffusion constant vs injection and temperature seems required for properly modelling the retrieved TRTS data. Nevertheless, our results highlight the strength of TRTS as a powerful non-contact method for analyzing the mobility in semiconductors as a function of key variables as the injection level and temperature.

## Acknowledgements

We acknowledge financial support from the grants PID2019-107808RA-I00 and TED2021-129624B-C44 funded by MCIN/AEI/10.13039/501100011033 and by "NextGenerationEU"/PRTR. We also acknowledge financial support from the Comunidad de Madrid through the projects 2017-T1/AMB-5207and 2021-5A/AMB-20942.